# MODELS OF LAMINATED PIEZOELECTRIC BEAMS INCLUDING EFFECT OF TRANSVERSE INTERACTIONS BETWEEN DIFFERENT LAYERS.

C. Maurini[*], F. dell'Isola[**] and J. Pouget[*]

[*] *Laboratoire d'Etudes Mécanique des Assemblages (FRE 2481), Université de Versailles/Saint-Quentin-en Yvelines, 45, ave. des Etats-Unis, 78035 Versailles cedex, France*

[**] *Dipartimento di Ingegneria Strutturale & Geotecnica, Università di Roma "La Sapienza", via Eudossiana 18, 00184 Roma, Italy*

*Summary* The present work attempts to present a consistent and efficient approach to piezoelectric laminated beams.. The influence of hypotheses on three-dimensional sectional deformations and stress distributions on the estimate of the beam electromechanical properties is analyzed. By exploiting a mixed variational formulation and Lagrange multipliers method, an Euler-Bernoulli-like beam model which accounts for transverse interactions between different is presented. The fully coupled electromechanical nature of the system is described by including both mechanical and electrical kinematical descriptors and both direct and inverse piezoelectric effects. For a sandwich piezoelectric beam and for a two-layers beam, expressions of the beam constitutive coefficients are provided and the main features of the proposed model are highlighted. Comparisons with experimental data and results from standard modelling approaches are presented. As main peculiarity, the proposed beam model coherently estimates the equivalent piezoelectric capacitance and transverse normal stress distribution also for beam composed by elastic and piezoelectric layers of comparable thickness.

Piezoelectric composites have been widely used for their sensor and actuator functions and research on this area opens many applications in the domain of adaptive structures and structural control [1]. Beam-shaped structures incorporating piezoelectric materials are often used in engineering applications as in robotic arms, airplane wings, helicopter rotor blades, positioning devices, etc. One-dimensional models of such structures provide analytical estimates of their electromechanical properties and useful tools for structural analysis and control system design [2].

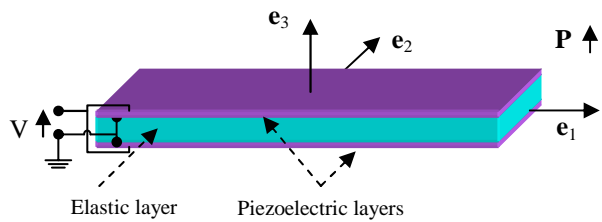

**Figure 1.** A laminated piezoelectric beam (sandwich configuration).

The aim of the present work is to discuss in details an issue which is often neglected in beam models of piezoelectric laminates: the influence of hypotheses on transverse normal stress ($T_{22}$) and strain ($S_{22}$) on the derivation of the constitutive properties of 1D models (mechanical stiffness, coupling coefficient and piezoelectric capacitance) from the 3D geometric and material data.

Beam models presented in literature (see e.g. [3,4]) usually retain classical assumptions for single-layered elastic structure by assuming that either transversal normal stress ($T_{22}$) or transversal normal strain ($S_{22}$) are vanishing layer by layer. However, as shown by finite element simulations based on a 3D model, these hypotheses are not verified in piezoelectric laminates with thickness-polarized piezoelectric layers. Indeed, when a potential difference is applied between the electrodes of a piezoelectric layer, it naturally tends to isotropically extend (or shrink) in the plane orthogonal to the polarization axis ($e_1$-$e_2$ plane). This behavior is in competition with Poisson effect in elastic layer: when one try to extend an elastic layer in one direction (say $e_1$), shrinking in the other direction (say $e_2$) is typically induced. When piezoelectric and elastic layers are bonded together to form a laminated piezoelectric beam, their contrasting behavior is conciliated by the appearance of non-negligible normal stresses both in the axial direction ($T_{11}$ stress) and transverse direction ($T_{22}$ stress). Such problem was discussed also by Beckert and Pfundtner [5], were classical assumptions on transverse stress are improved and complex interactions between different layers in composite piezoelectric laminates are underlined. However, only actuation functions are considered and no attention is paid to the determination of purely electrical properties such as the equivalent piezoelectric capacitance.

In the present paper, an efficient model of laminated piezoelectric beams including transverse interactions between different layers is proposed. It includes both sensory and actuation effects and is based on the following hypotheses:

i) equivalent single-layer Euler-Bernoulli kinematics (only axial extensional strain $S_{11}$ with a linear distribution along the thickness);
ii) layerwise linear distribution of mechanical stress with non-vanishing transverse normal stress (i.e. $T_{22}$);
iii) layerwise linear distribution of electric potential;
iv) layerwise constant distribution of the electric displacement.

The transverse normal stress $T_{22}$ are not determined by corresponding equilibrium equations since transversal deformations are not included in the Euler-Bernoulli kinematics and no energy contribution is associated to them. Their distribution in the different layers is found by imposing the following additional conditions (null force resultant and null moment resultant):

$$N_2 = \int_{Section} T_{22} = 0, \quad M_2 = \int_{Section} z T_{22} = 0. \qquad (1)$$

The proposed model which is characterized by assuming null transverse stress resultants (NSR model) is compared to the models assuming either pointwise null transverse stress ($T_{22}=0$, NS model) or null transverse sectional deformations ($S_{22}=0$, ND model).

The model above imposes at the same time hypotheses on the distribution of electromechanical kinematical and dynamical state field (strain, electric potential, stress, and electric displacement). Hence, in order to rationally deduce the beam models from the three-dimensional description, a mixed variational formulation based on the Hellinger-Prange-Reissner functional for piezoelectricity is adopted. Moreover, the additional conditions on transverse stresses (1) are introduced into the variational formulation through Lagrange multipliers method.

The influence of the introduced stress field $T_{22}$ and the associated weak conditions on the beam constitutive equations is discussed in details. In particular, it is shown that the different hypotheses (null transverse stress resultants, null transverse stress, null transverse strain) on transversal normal stress ($T_{22}$) and strain ($S_{22}$) strongly influence the estimate for the equivalent piezoelectric capacitance. Comparisons to experimental data and finite element simulations are reported (see Table 1) and the peculiar ability of the proposed model of correctly estimating the equivalent piezoelectric capacitance and the transverse normal stress distribution is shown.

|  | ND model | NS model | NSR model | Experimental |
|---|---|---|---|---|
| Capacitance per unit line | 2.09 nF/mm | 3.62 nF/mm | 2.81 nF/mm | 2.86 nF/mm |
| Error (%) | +26.7% | -26.6% | -1.75 % | ---- |

**Table 1.** Comparison between piezoelectric capacitance per unit line predicted by the proposed model (NSR model), standard models (ND and NS models) and the one measured experimentally. The values refer to a sandwich piezoelectric beam (see Figure 1) made of a central aluminium core on which two piezoelectric layers (PZT-5H) are symmetrically bonded (beam width 17.8mm, piezoelectric layers thickness 0.27mm, aluminium layer thickness 2.0 mm).